\documentclass[12pt,a4paper]{iopart}
\usepackage{iopams}
\usepackage[english]{babel}
\usepackage{float}
\usepackage[dvips]{graphicx}
\usepackage{rotating}
\usepackage{epsfig}
\usepackage[dvips]{color}
\definecolor{Red}{named}{Red}

\begin{document}
\title[The Effect of Thermal Neutrino Motion on the Matter Power Spectrum]{The
Effect of Thermal Neutrino Motion on the Non-linear Cosmological Matter Power Spectrum}
\author{Jacob Brandbyge$^1$, Steen Hannestad$^1$, Troels Haugb{\o}lle$^{1,2}$, Bjarne
  Thomsen$^1$}
\ead{jacobb@phys.au.dk, sth@phys.au.dk, haugboel@phys.au.dk, bt@phys.au.dk}
\address{$^1$Department of Physics and Astronomy, University
of Aarhus, Ny Munkegade, Bygn. 1520, 8000, Aarhus, Denmark
\\$^2$Instituto de F\'{\i}sica Te\'{o}rica UAM-CSIC, Universidad Aut\'{o}noma
de Madrid, Cantoblanco, 28049, Madrid, Spain}
\date{\today}
\begin{abstract}
We have performed detailed studies of non-linear structure formation
in cosmological models with light neutrinos. For the first time the
effect of neutrino thermal velocities has been included in a
consistent way, and the effect on the matter power spectrum is found
to be significant. The effect is large enough to be measured in
future, high precision surveys. Additionally, we provide a simple
but accurate analytic expression for the suppression of fluctuation
power due to massive neutrinos. Finally, we describe a simple and
fast method for including the effect of massive neutrinos in
large-scale $N$-body simulations which is accurate at the 1\%
level for $\sum m_\nu \lesssim 0.15 \,{\rm eV}$.
\end{abstract}
\pacs{98.65.Dx, 95.35.+d, 14.60.Pq\hspace{\stretch{1}} Preprint:
IFT-UAM/CSIC-08-11} \maketitle
\section{Introduction}
Neutrinos are among the most abundant particles in our Universe and
therefore play an important role in the formation of large-scale structure. In
the past 10 years neutrino oscillation experiments have verified
that neutrinos have small, but non-zero masses and that they
therefore contribute to the dark matter density in the Universe.

Neutrino oscillation experiments have established the existence of
two distinct neutrino mass splittings,
$\Delta{m}^2_{ij}=m^2_i-m^2_j$. $\Delta{m}^2_{12}\simeq 7 \times
10^{-5} \, {\rm eV}^2$ has been found by observing Solar and reactor
neutrinos, and $|\Delta{m}^2_{23}| \simeq 2.5 \times 10^{-3} \, {\rm
eV}^2$ by observing atmospheric neutrinos and by accelerator
experiments (see e.g.\ \cite{Fogli:2006yq,Maltoni:2004ei} for recent
data analyses). This gives a lower bound on the sum of the neutrino
masses which is $\sum m_\nu \gtrsim 0.05 \, {\rm eV}$ for the normal
hierarchy and $\sum m_\nu \gtrsim 0.1 \, {\rm eV}$ for the inverted
hierarchy.

Using a combination of cosmological observables such as the cosmic
microwave background, large scale structure, and type Ia supernovae
a bound on the sum of neutrino masses of $\sum m_\nu \lesssim
0.2-0.7$ eV (95\% C.L.) has been derived \cite{Zunckel:2006mt,%
Cirelli:2006kt,Goobar:2006xz,Kristiansen:2006xu,%
Seljak:2006bg,Hannestad:2003xv,Hannestad:2006zg,Hannestad:2007tu}.
The actual value of the bound depends both on the specific
combination of data sets used and on the cosmological model. At the
moment a fairly robust upper limit can be taken to be roughly $\sum
m_\nu \lesssim 0.6-0.7$ eV.

This bound mainly uses structure formation data in the linear regime
($k \lesssim 0.15 \, h\, {\rm Mpc^{-1}}$), but even on such large scales some
non-linear contamination is present and must be modelled. This has
been done in detail for variations of the standard $\Lambda$CDM
model, but so far not in a truly consistent way for models which
include massive neutrinos.

For the study of neutrino mass bounds with present data the current
precision of theoretical power spectrum calculations in the regime
$k \sim 0.1-1 \, h \, {\rm Mpc}^{-1}$ is sufficient in the sense that the
final result does not depend very significantly on the method used
(however, see e.g.\ \cite{Hannestad:2007dd} for an exception).
However, future high precision lensing and galaxy redshift surveys,
such as the LSST \cite{LSST}, will constrain the matter power spectrum to percent level precision
on scales of $k \sim 0.1-1 \, h \, {\rm Mpc}^{-1}$.

At this level of accuracy the inclusion of neutrinos in the
calculation of the matter power spectrum will be crucial, even for
neutrino masses close to the current lower bound. While the effect
of neutrinos in the linear theory power spectrum is understood in
detail much work remains to be done in the semi-linear and
non-linear regimes. Furthermore, semi-analytic models of non-linear structure
formation such as the halo model must always be checked against {\it
N}-body simulations.

Neutrinos have thermal velocities which exceed typical gravitational flow velocities by an order of magnitude or more, even at low redshift. Therefore, when setting up initial conditions for {\it N}-body simulations the neutrino thermal velocities need to be added to the gravitational flow velocities found from the neutrino transfer function. Including neutrino thermal velocities in {\it N}-body simulations is not without problems because the thermal motion introduces a noise term
which will affect the underlying gravitational evolution unless high
resolution is used.

The problem is similar to the well known problem that particles in
an {\it N}-body simulation must be put in a configuration like a grid or a
glass which respects small-scale homogeneity in order not to
introduce a large Poisson noise term. When thermal velocities are
included momentum phase space must be sampled and the
initial thermal velocities should preferably be chosen in a way where small-scale
momentum conservation is enforced.

In the present paper we present the first {\it N}-body simulations
in which neutrino thermal velocities have been included and
convergent results achieved for $k < 1.5 \, h\,{\rm Mpc^{-1}}$. In the early 90s \cite{Klypin:1992sf,Primack:1994pe} included neutrino thermal velocities albeit for a model with much higher neutrino masses, and no convergent results were proven. We also note
another recent paper \cite{Colin:2007bk} in which thermal motion has
been included for warm dark matter, a set-up somewhat different from
what is studied here. Neutrino clustering in the small-scale
limit has been studied using a completely different approach
\cite{Singh:2002de,Ringwald:2004np} where the full Boltzmann
equation has been solved for neutrinos in a background given by pure
cold dark matter (CDM), i.e.~no feed-back is included and neutrinos are treated as
tracer particles in the underlying gravitational potential. This
method allows for arbitrarily high resolution, but is not truly
self-consistent.

In Section \ref{secIC} we describe the linear theory evolution and
the set-up of initial conditions for our neutrino {\it N}-body
simulations. In Section \ref{secNbody} we describe how our {\it
N}-body simulations are performed, and in Section \ref{secResults}
the results are described in detail. Finally, Section
\ref{secDiscussion} contains a discussion and our conclusions.

\section{Linear Evolution of Perturbations and Initial Conditions}\label{secIC}
\subsection{Linear Theory}

Evolving the primordial density perturbations set down by inflation
to the present involves several steps. As long as perturbations
remain small the evolution can be calculated precisely using the
linearised Einstein and Boltzmann equations \cite{Ma}, using
software such as CAMB \cite{CAMB} or CMBFAST \cite{CMBFAST}.

In this regime the power spectrum can be factorised into a
primordial component, $P_0(k)$, and a transfer function (TF), $T(k,z)$, which
contains all information about the evolution of structure so that
$P(k,z) = P_0(k) T^2(k,z)$.

However, once structure enters the non-linear regime precision
studies require the use of {\it N}-body simulations. To set up the initial
conditions (ICs) for our simulations we have calculated the
TFs using CMBFAST (using CAMB yields similar results).

We have assumed a flat cosmological model with density parameters $\Omega_{b}=0.05$,
$\Omega_{m}=0.30$ and $\Omega_\Lambda=0.70$ for the baryon, matter and
cosmological constant components, respectively, and a Hubble parameter
of $h=0.70$. We vary the CDM and neutrino density parameters
($\Omega_{\rm CDM}$ and $\Omega_\nu$, respectively) fulfilling the
condition $\Omega_{\rm CDM}+\Omega_\nu = 0.25$. We have assumed
a primordial power spectrum of the form $P_0(k) = A k^{n}$ with
$n=1$, i.e.\ a standard scale-invariant Harrison-Zel'dovich
spectrum. The amplitude, giving $\sigma_8=0.89$ for a pure $\Lambda {\rm CDM}$ model, is chosen so as to fit the CMBR \cite{spergel1} on large scales.

\begin{figure}
\begin{center}
  \includegraphics[width=0.6\textwidth]{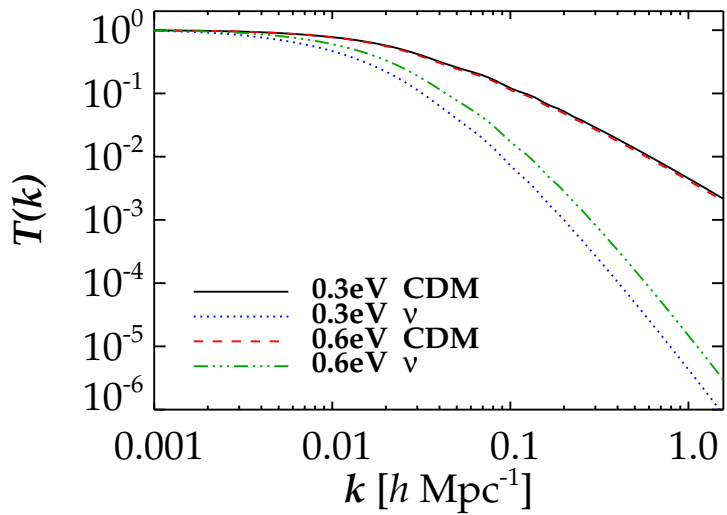}
  \caption{The linear theory transfer functions at $z=4$ for the CDM
  and neutrino components.}
  \label{fig:TF02}
\end{center}
\end{figure}

Different TFs at a redshift of $4$ are shown in Fig.~\ref{fig:TF02}.
The effect of neutrino free-streaming on the TF is
clearly seen, and is more pronounced in the lower mass neutrino
case.

\subsection{Initial Conditions with Two Species}

\begin{figure*}[ht]
\begin{center}
   \includegraphics[width=1.0\textwidth]{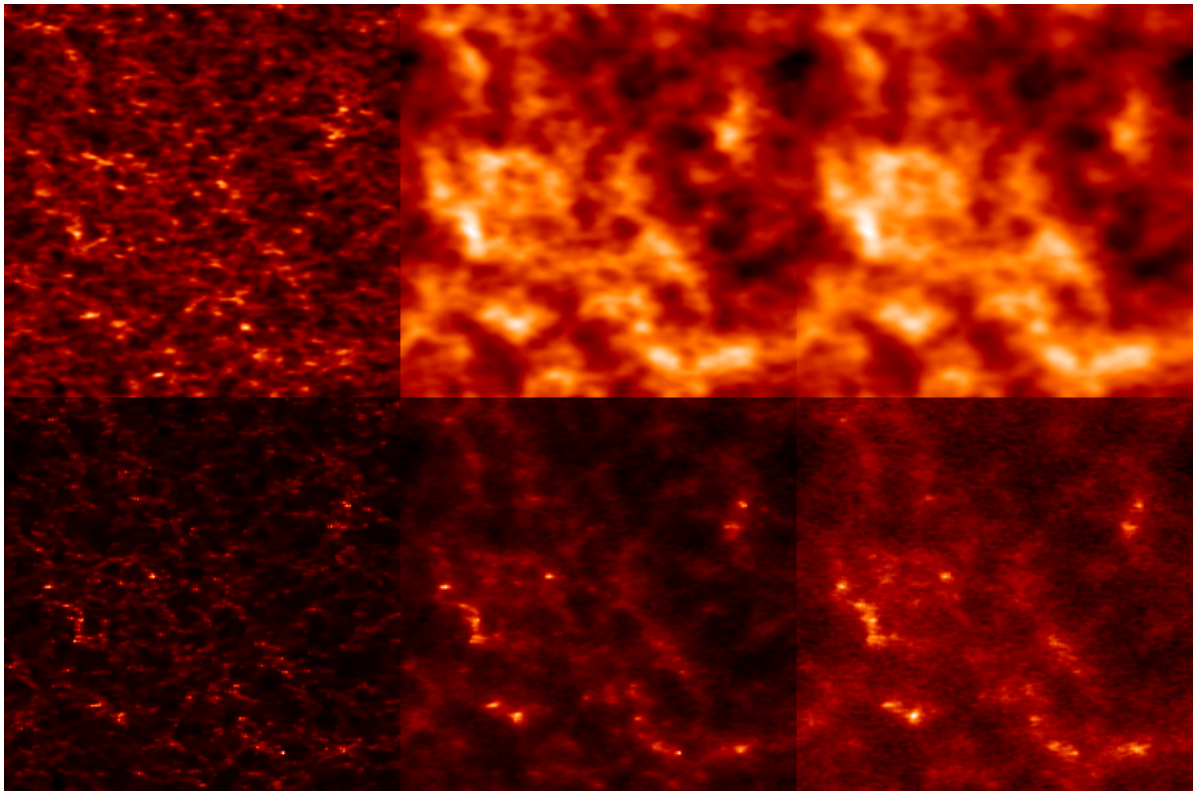}
   \caption{Images of the CDM and neutrino density distributions in a
   slice of the simulation volume. The images span $512\,h^{-1} \, {\rm Mpc}$ on a
   side and has a depth of $10\,h^{-1}\, {\rm Mpc}$. To produce the images we have
   interpolated the masses of the {\it N}-body particles to a regular grid with the adaptive smoothing kernel of
   \cite{monaghan}. The images show the densities for the CDM
   component (left), neutrinos with $\sum m_\nu=0.6 \, {\rm eV}$ (middle), and neutrinos with
   $\sum m_\nu=0.3 \, {\rm eV}$ (right). The top row is at $z_i=4$ and the bottom
   row at $z=0$. To enhance the dynamic range of the CDM structures
   the square root has been taken of the CDM density field in the
   $z=0$ image. The $\sum m_\nu=0.3 \, {\rm eV}$ neutrino image at $z=0$ displays
   artificial small-scale structures in the voids caused by neutrino
   $N$-body particle shot-noise. All the images are made from
   simulations with $512^3$ neutrino $N$-body particles.}
   \label{fig:rho}
\end{center}
\end{figure*}

The TFs are used to generate the position and velocity ICs for the {\it N}-body particles. The Zel'dovich
Approximation (ZA) \cite{Zeldovich:1969sb}, based on first-order
Lagrangian perturbation theory, is a standard way to calculate the
ICs. Using second-order Lagrangian perturbation theory (2LPT)
\cite{Scoccimarro1,Bouchet} it is possible to generate ICs in the
quasi-linear regime. We are particularly interested in these
second-order corrections because the presence of neutrino thermal
velocity noise requires simulations to be started at low redshift,
close to the non-linear regime.

We have used a modified version of the 2LPT initial conditions code of
\cite{Crocce} to generate the ICs for our simulations. The
modifications are described in detail below.

In the {\it N}-body simulations we include two particle species, CDM and
neutrinos. It is not possible to generate particle ICs for each
species simultaneously
with the ZA+2LPT formalism. Instead we generate ICs one
particle species at a time by using their respective TFs (for CDM we
have used a weighted sum of the CDM and baryon TFs for consistency, even though this
only matters at high redshift). Thereafter, the {\it N}-body particle masses for each
species are scaled so that they make up their proper fraction of
$\Omega_{m}$.

The initial positions for the {\it N}-body particles are found by adding a
displacement field to a regular grid. A standard way to get the
particle velocities is to differentiate the initial position
displacements. This procedure involves using several numerically
determined fitting factors, and therefore breaks down when two
species with different TFs are present since then the growth factor
is both species and mode dependent. Instead, we get the velocities
by generating two displacement fields centered around our starting
redshift and then take the time difference. We have tested that
these velocities do not depend on the distance in redshift between
the two extra displacement fields in a suitable range around our
starting redshift.

2LPT involves a relation between the first- and second-order growth
factors. But since the perturbed energy density even at $z_i=4$
($z_i$ designates the $N$-body starting redshift) is vastly dominated
by CDM, we can neglect the
neutrino contribution to the driving term for the CDM growth factor
since this would give a small correction to a second-order term.
Because the neutrinos are smoothly distributed, first-order theory
is very accurate for them, and therefore the second-order
corrections give negligible contributions to the neutrinos. We note
that this is somewhat similar to the approach taken in
Refs.~\cite{Hannestad:2005bt,Hannestad:2006as,saito1} for the case
of semi-analytic power spectrum modelling.

In the top row of Fig.~\ref{fig:rho} the CDM and neutrino distributions are shown
at $z_i=4$. Since we run simulations with different numbers of CDM and
neutrino $N$-body particles, the amplitude and
phase of the common large-scale
modes of each species have been generated with the same random
numbers. Therefore,
from Fig.~\ref{fig:rho} one can identify the components as having
the same large-scale structure. Because of the higher value of the
CDM TF at small scales, this component has additional clearly
visible small-scale structure. As expected, the neutrino distributions are more
homogeneous than their CDM counterpart. The $\sum m_\nu=0.6 \, {\rm
eV}$ and $\sum m_\nu=0.3 \, {\rm eV}$ distributions are similar at $z_i=4$,
with the latter distribution being marginally more homogeneous, as can
also be inferred from the TFs in Fig.~\ref{fig:TF02}.

\subsection{Thermal Velocities}\label{sec_vt}

\begin{figure}
\begin{center}
  \includegraphics[width=0.6\textwidth]{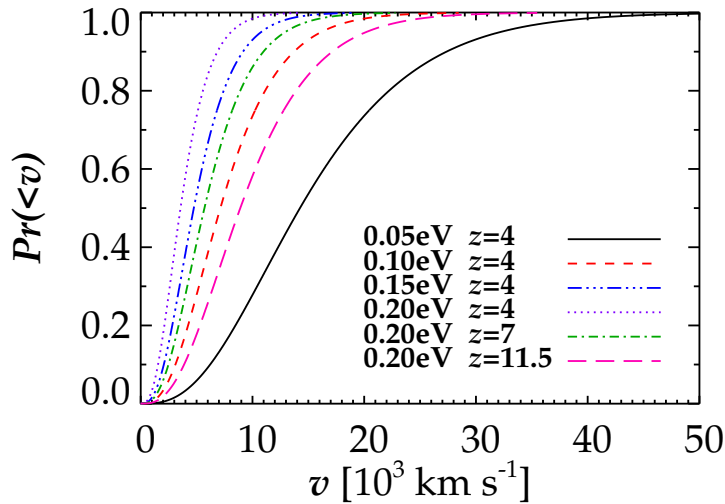}
  \caption{The cumulative Fermi-Dirac distributions as a function of
  velocity for most of the simulations listed in Table \ref{fig:table1}.}
  \label{fig:FD}
\end{center}
\end{figure}

Since the neutrino mass is small the thermal velocities
cannot be neglected even at low redshift if percent level
precision is to be obtained. Instead, the velocity of a given
neutrino $N$-body particle contains a thermal component drawn from an equilibrium
Fermi-Dirac (FD) distribution. Assuming isotropy for the thermal
component the probability, $Pr$, for a neutrino having a momentum
smaller than $p$ is given by
\begin{equation}
{Pr}(<p)=N\int_0^p{\frac{{p'}^2}{e^{p'c/k_bT_\nu}+1}dp'},
\label{eq:FDeq}
\end{equation}
where $N$ is a normalisation which ensures that the probability is
bounded between $0$ and $1$. Fig.~\ref{fig:FD} shows the cumulative
FD distributions for different one-particle neutrino
masses. The magnitudes of the neutrino thermal velocities were drawn
randomly from the FD velocity distribution, and the directions of the
thermal velocities were drawn at random as well. This reflects the fact
that the neutrino velocity can be split into two uncorrelated contributions. An
equilibrium, random contribution arising from the FD distribution
and an out-of-equilibrium gravitational flow contribution found from the ZA+2LPT
formalism.

The neutrinos decoupled from the baryon-photon plasma before the
annihilation of electrons and positrons, and as a result the
neutrino temperature is related to the photon temperature by the
approximate relation $T_\nu \simeq T_\gamma({4}/{11})^{1/3}$.
Assuming $p\ll{m}_\nu$ at late times the 3 neutrino generations
contribute the following to the present neutrino density parameter
\cite{Dicus:1982bz,Dodelson:1992km,%
Hannestad:1995rs,Dolgov:1997mb,%
Steigman:2001px,Mangano:2001iu,osc,%
Mangano:2005cc}
\begin{equation}
\Omega_{\nu,0} =\frac{\sum{m_\nu}}{93.8 h^2 \, {\rm eV}}.
\label{eq:Omega}
\end{equation}

The finite number of neutrino {\it N}-body particles gives
rise to noise in the power spectrum. Physically this leads to
spuriously strong clustering of neutrinos on small scales, even at
early times. As will be discussed below we have carefully checked
that this noise term is negligible on the scales of interest in our
simulations.

However, since the neutrino thermal velocity distribution redshifts
as $a^{-1}$, and the neutrino and CDM out-of-equilibrium density
contrasts and velocity fields increase with the expansion of the
Universe, it is paramount to start the {\it N}-body simulations as late,
and therefore as close to the non-linear regime, as possible. This in
turn constrains the scales on which clustering can be
reliably calculated.

We have tested the validity of starting the {\it N}-body simulations
as late as $z_i=4$. This has been done by starting 4 pure $\Lambda$CDM simulations, at
$z_i=$ 49, 11.5, 7 and 4, and evolving them to the present. The difference in
the matter power spectrum between these simulations is at the few
percent level on small scales. Since the neutrino simulations
include a smaller amount of CDM this is an upper limit. The 2LPT
corrections were crucial for achieving such a small discrepancy.
Since we are interested in quantifying the relative effect of
including neutrinos, and {\it not} the exact absolute value of the
matter power spectrum, we are justified in choosing $z_i=4$ as the
lowest starting redshift for our {\it N}-body simulations. If the
absolute power spectrum is desired at the percent level a higher
starting redshift should be chosen, but we stress that this will
have a minimal effect on the relative change in the matter power
spectrum coming from including the neutrino component.

\begin{table*}[t]
{\footnotesize 
  \hspace*{-0.8cm}\begin{tabular}
  {|c||c@{\hspace{2pt}}c@{\hspace{2pt}}c@{\hspace{2pt}}c|c@{\hspace{2pt}}c@{\hspace{2pt}}c@{\hspace{2pt}}c@{\hspace{2pt}}c@{\hspace{2pt}}c|c@{\hspace{2pt}}c@{\hspace{2pt}}c@{\hspace{2pt}}c@{\hspace{2pt}}c|c@{\hspace{2pt}}c@{\hspace{2pt}}c|}                      \hline
  & $S^0_{1}$& $S^0_{2}$&$S^0_{3}$&$S^0_{4}$ & $S^{0.15}_{1}$ & $S^{0.15}_{2}$ &  $S^{0.15}_{3}$ & $S^{0.15}_{4}$ & $S^{0.15}_{5}$ & $S^{0.15}_{6}$ &$S^{0.3}_{1}$ & $S^{0.3}_{2}$ &  $S^{0.3}_{3}$ & $S^{0.3}_{4}$ & $S^{0.3}_{5}$& $S^{0.45}_{1}$ &$S^{0.45}_{2}$&$S^{0.45}_{3}$\\       \hline\hline
$N_{\nu}$                      & 0 &0 & 0 &0    & ${256}^3$   & 0    & $128^3$ & $256^3$& ${256}^3$   & 0& ${256}^3$   & 0         & $128^3$ & $256^3$&    ${512}^3$          & ${256}^3$   &0&$256^3$\\
$\sum m_\nu~[{\rm eV}]$                   & 0 &0& 0 &0    & 0.15&0.15&0.15&0.15&0.15&0.15&0.3         &    0.3 &  0.3     &0.3  & 0.3        & 0.45         &0.45& 0.45 \\
$\Omega_{\nu,0}$~[$\%$]            & 0  &0& 0  &0   & 0.33&0.33&0.33&0.33& 0.33&0.33&0.65      & 0.65 & 0.65&0.65&0.65          & 0.98 &0.98 & 0.98\\
FD               & No&No& No&No   & No       & No      & Yes       &Yes& No       & No& No       & No      & Yes       &Yes & Yes      & No        &No& Yes \\
$z_i$                          & 4 &7&11.5&49   & 4&4&4&4&49&49&4           & 4         &  4      &4  & 4                     & 4     &4&4\\\hline
  \end{tabular}
  \hspace*{-0.8cm}\begin{tabular}{|c||c@{\hspace{2pt}}c@{\hspace{2pt}}c@{\hspace{2pt}}c@{\hspace{2pt}}c@{\hspace{2pt}}c@{\hspace{2pt}}c@{\hspace{2pt}}c|c@{\hspace{2pt}}c@{\hspace{2pt}}c@{\hspace{2pt}}c|c@{\hspace{2pt}}c@{\hspace{2pt}}c@{\hspace{2pt}}c|c@{\hspace{2pt}}c@{\hspace{2pt}}c|}                      \hline
  &$S^{0.6}_{1}$ &$S^{0.6}_{2}$&$S^{0.6}_{3}$&$S^{0.6}_{4}$& $S^{0.6}_{5}$ &$S^{0.6}_{6}$ & $S^{0.6}_{7}$ & $S^{0.6}_{8}$&$S^{0.6}_{9}$ &$S^{0.6}_{10}$&$S^{0.6}_{11}$&$S^{0.6}_{12}$&$S^{0.6}_{13}$ &$S^{0.6}_{14}$&$S^{0.6}_{15}$&$S^{0.6}_{16}$ &$S^{0.6}_{17}$&$S^{0.6}_{18}$&$S^{0.6}_{19}$   \\       \hline\hline
$N_{\nu}$                    &   ${256}^3$   &0&$32^3$ &$64^3$ & $128^3$ &$256^3$& ${512}^3$ & ${1024}^3$&   ${256}^3$   &0&$128^3$ &$256^3$&   ${256}^3$   &0&$128^3$ &$256^3$ & ${128}^3$ &  $256^3$ &$512^3$ \\
$\sum m_\nu~[{\rm eV}]$           & 0.6        &0.6& 0.6     &0.6  & 0.6  &0.6 & 0.6 & 0.6& 0.6  &0.6  & 0.6 & 0.6& 0.6  &0.6  & 0.6 & 0.6 &0.6  & 0.6 & 0.6\\
$\Omega_{\nu,0}$~[$\%$]            & 1.3      &1.3& 1.3  &1.3  & 1.3&1.3 &1.3 & 1.3& 1.3&1.3  & 1.3& 1.3& 1.3&1.3  & 1.3& 1.3&1.3  & 1.3& 1.3\\
FD               & No        &No& Yes       &Yes & Yes&Yes & Yes& Yes& No        &No& Yes       &Yes & No        &No& Yes       &Yes &Yes& Yes       &Yes \\
$z_i$                          & 4           &4 &  4       & 4 &4   &4&4&4& 7    &7&7&7 & 11.5    &11.5&11.5&11.5  & 49 &49   &49\\\hline
  \end{tabular}
  }
  \caption{Parameters for our {\it N}-body simulations. $N_\nu$ is the number of neutrino
  $N$-body particles. $\sum m_\nu $ is the total neutrino mass, and it is in
  all cases related to the one-particle neutrino mass, $m_\nu$, by $\sum m_\nu = 3 m_\nu$. $\Omega_{\nu,0}$ is the
  fraction of the critical density contributed by the neutrinos today
  (see Eq.~\ref{eq:Omega}), and FD indicates whether or not the
  neutrinos have been given a thermal velocity from the relativistic
  Fermi-Dirac distribution (see Eq.~\ref{eq:FDeq}). Finally, $z_i$ indicates the $N$-body
  starting redshift. Note that $N_\nu=0$ combined with $\sum m_\nu > 0$ indicates that the neutrino distribution has been kept totally homogeneous in the $N$-body simulation.}
  \label{fig:table1}
\end{table*}

\section{{\emph N}-body Simulations}\label{secNbody}
The {\it N}-body simulations were performed with the publicly available
{\it N}-body code \textsc{gadget-2} \cite{Springel2} run in the hybrid
TreePM mode. Gas physics was neglected, since it does not
significantly affect the scales of interest.

The {\it N}-body simulations are evolved with Newtonian dynamics. There
are several reasons for this. First, the ZA+2LPT formalism is stricly
Newtonian. Second, even though the neutrino thermal velocities
approach up to 15\% of the speed of light, the relativistic corrections are much
smaller than the desired accuracy.

The simulations are listed in Table \ref{fig:table1}, and include
$256^3$ CDM particles in a $512h^{-1} \, {\rm Mpc}$ box. By running a simulation with $512^3$ CDM particles we have tested the validity of using $256^3$ CDM particles. The simulation with the highest number of $N$-body particles gave a correction to the matter power spectrum which was at the one percent level on small scales. Since we only quantify the relative effect of including neutrinos, $256^3$ CDM particles is sufficient. To test the convergence of the matter and neutrino power spectra we have made
runs with different numbers of neutrino $N$-body particles.

As a reference, and in order to test how the neutrino thermal velocity
affect the power spectra, we have run simulations with $256^3$ neutrinos
with no thermal velocities included.

We have also tested how much the neutrinos contribute to the matter
power spectrum by running simulations with the neutrinos included in
the linear evolution but neglecting the perturbed neutrino component
in the {\it N}-body simulations. Here, the neutrinos were still
included in the calculation of the Hubble parameter so that the
background evolution is identical. Physically, this corresponds to
keeping the neutrino distribution totally homogeneous in the {\it
N}-body simulation.

The gravitational Tree part in \textsc{gadget}-2 sets the timestep
according to the gravitational acceleration, not the neutrino
thermal velocity, so that the fast moving neutrinos may not be
accurately evolved. We have tested if the neutrino timestep was
small enough to simulate the scales of interest by decreasing it by roughly a factor of 5 in the Tree
part. The correction found was well below $0.1\%$. This very small correction can be explained by the fact that the neutrino small-scale structure does not contribute to the matter power spectrum on the smallest scales simulated.

Finally, we have investigated the effect of the finite box size for the case with $\sum m_\nu=0.6 \, {\rm eV}$ and $z_i=4$. This has been done by reducing the size of the simulation volume as well as the number of CDM particles by a factor of two. A different set of random numbers was used to generate the initial conditions. The amount of suppresion due to neutrinos and the position of the turnover in the difference power spectrum were found to be consistent with the results presented in Fig.~\ref{fig:damping}.

\section{Results}\label{secResults}

\subsection{Damping and Convergence of the Power Spectrum}

\begin{figure}[t]
\begin{center}
  \includegraphics[width=\textwidth]{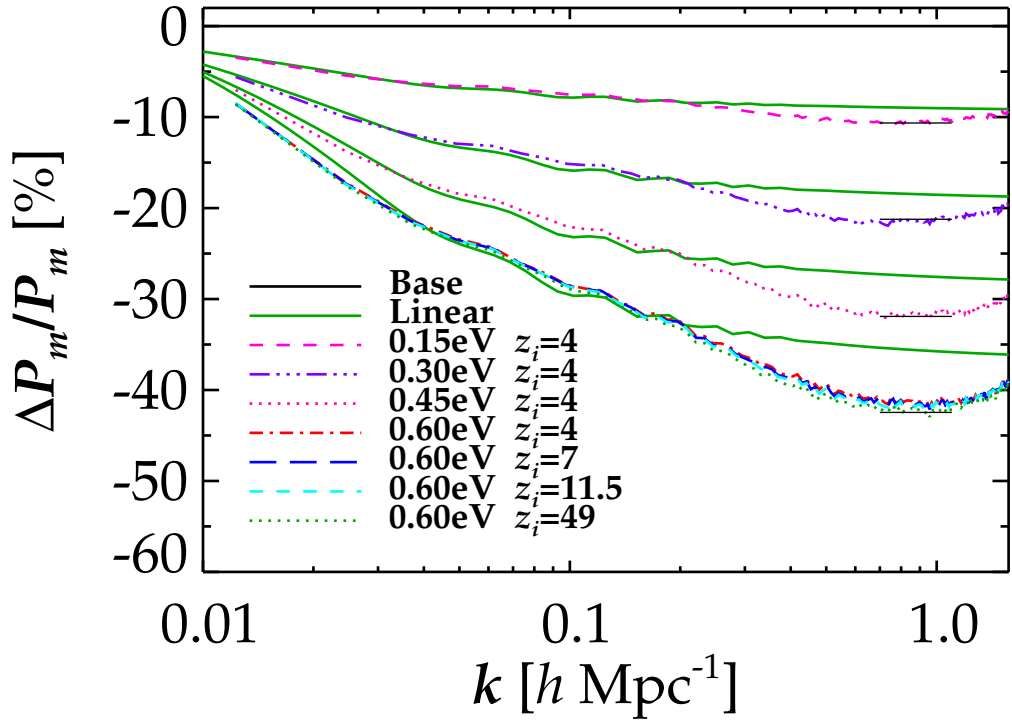}
    \caption{Relative differences in
    the matter power spectra at $z=0$ between pure $\Lambda$CDM
    models and models with neutrinos included. The
    differences expected from linear theory are also shown. The horizontal black lines indicate a relative power spectrum suppression of $-9.8 \, \Omega_\nu / \Omega_m$.}
    \label{fig:damping}
\end{center}
\end{figure}

Fig.~\ref{fig:damping} shows the well-known scale dependent damping
of power when neutrinos are included. As expected, the
damping is much larger in the $\sum m_\nu=0.6 \, {\rm eV}$ case as
compared to $\sum m_\nu=0.15 \, {\rm eV}$ because $\Omega_{\rm CDM}$
is smaller in the former case even though the neutrino thermal
velocity is largest for the lower mass neutrino.

From Fig.~\ref{fig:damping} it can be seen that in the $\sum
m_\nu=0.6 \, {\rm eV}$ case the damping is almost independent of $z_i$. Note
that the relative decrease in power for a given simulation with
neutrinos is taken with respect to a $\Lambda$CDM simulation started
at the same $z_i$. This has been done to remove the dependence on the
starting redshift when comparing the models.

Fig.~\ref{fig:damping} also shows the damping expected from linear
theory, which is accurate out to $k \sim 0.2 h \, {\rm Mpc}^{-1}$.
On smaller scales non-linear theory predicts a substantially larger
damping of the power spectrum. The departure from linear theory
increases with higher $\sum m_\nu$. Note that non-linear theory
predicts a turnover in the difference power spectrum. At the starting
redshift, $z_i$, this turnover is not present, neither in the TFs nor
when the ICs have
been calculated. 

From a mode-coupling point
of view the turnover occurs because the small-scale modes ($k \sim 1 \, h\, {\rm
Mpc^{-1}}$) in the simulations with neutrinos get relatively more
out of their coupling to the large-scale modes than do the same
modes in the pure $\Lambda {\rm CDM}$ simulations. The turnover does not appear because of mode-coupling between the small-scale modes themselves. Physically, large-scale potential gradients are needed to make the small-scale structures grow substantially, and the large-scale perturbations are less affected by neutrino free-streaming than are the small-scale perturbations. The non-linear collapse of the halos then gives mode-coupling and, combined with the characteristic damping of the difference power spectrum caused by neutrino free-streaming, gives the turnover.

\begin{figure}
   \noindent
   \begin{minipage}{0.49\linewidth}
      \hspace*{-1.1cm}\includegraphics[width=1.2\linewidth]{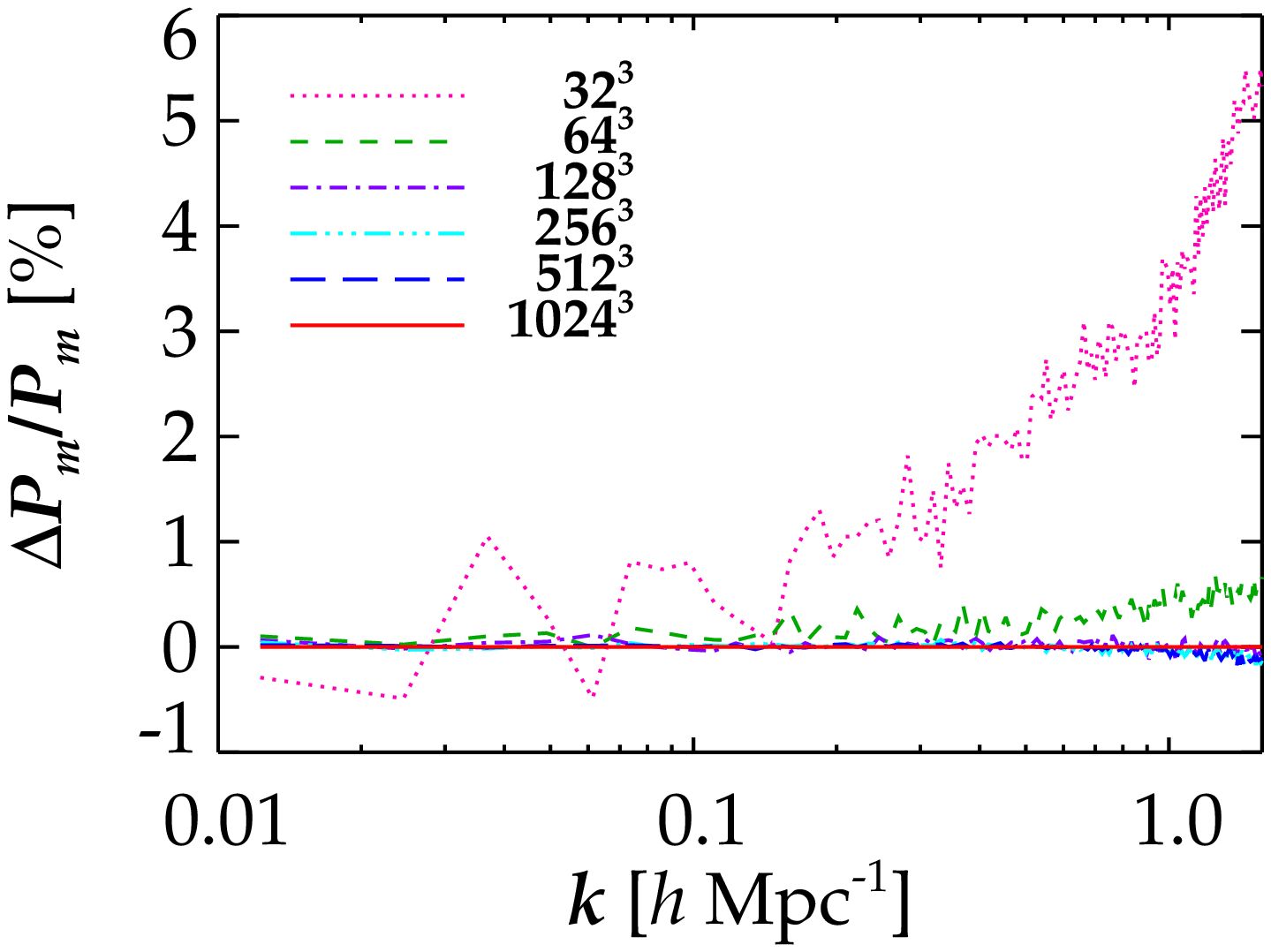}
   \end{minipage}
   \begin{minipage}{0.49\linewidth}
      \hspace*{-0.5cm}\includegraphics[width=1.2\linewidth]{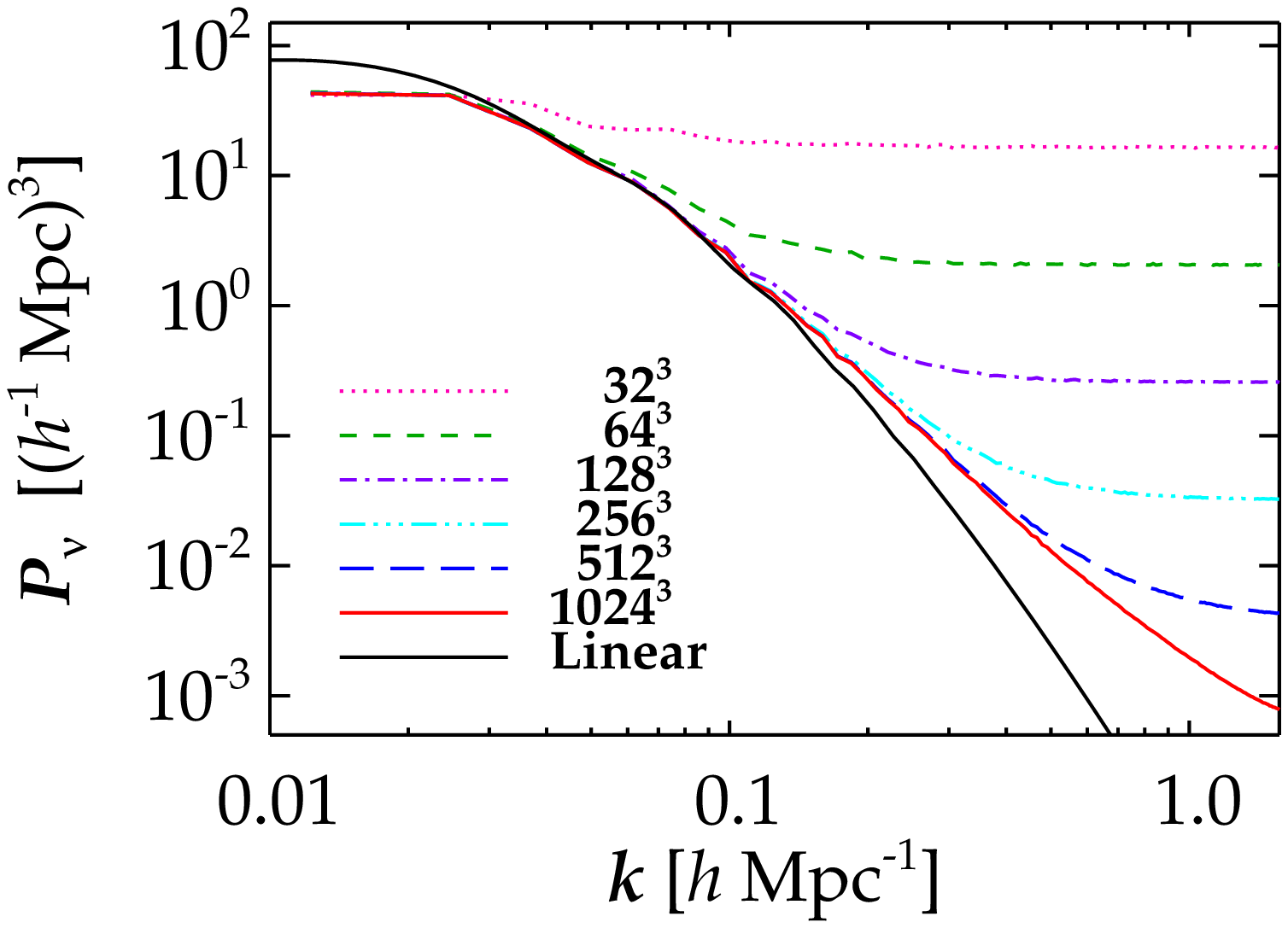}
   \end{minipage}
%
   \noindent
   \begin{minipage}{0.49\linewidth}
      \hspace*{-1.1cm}\includegraphics[width=1.2\linewidth]{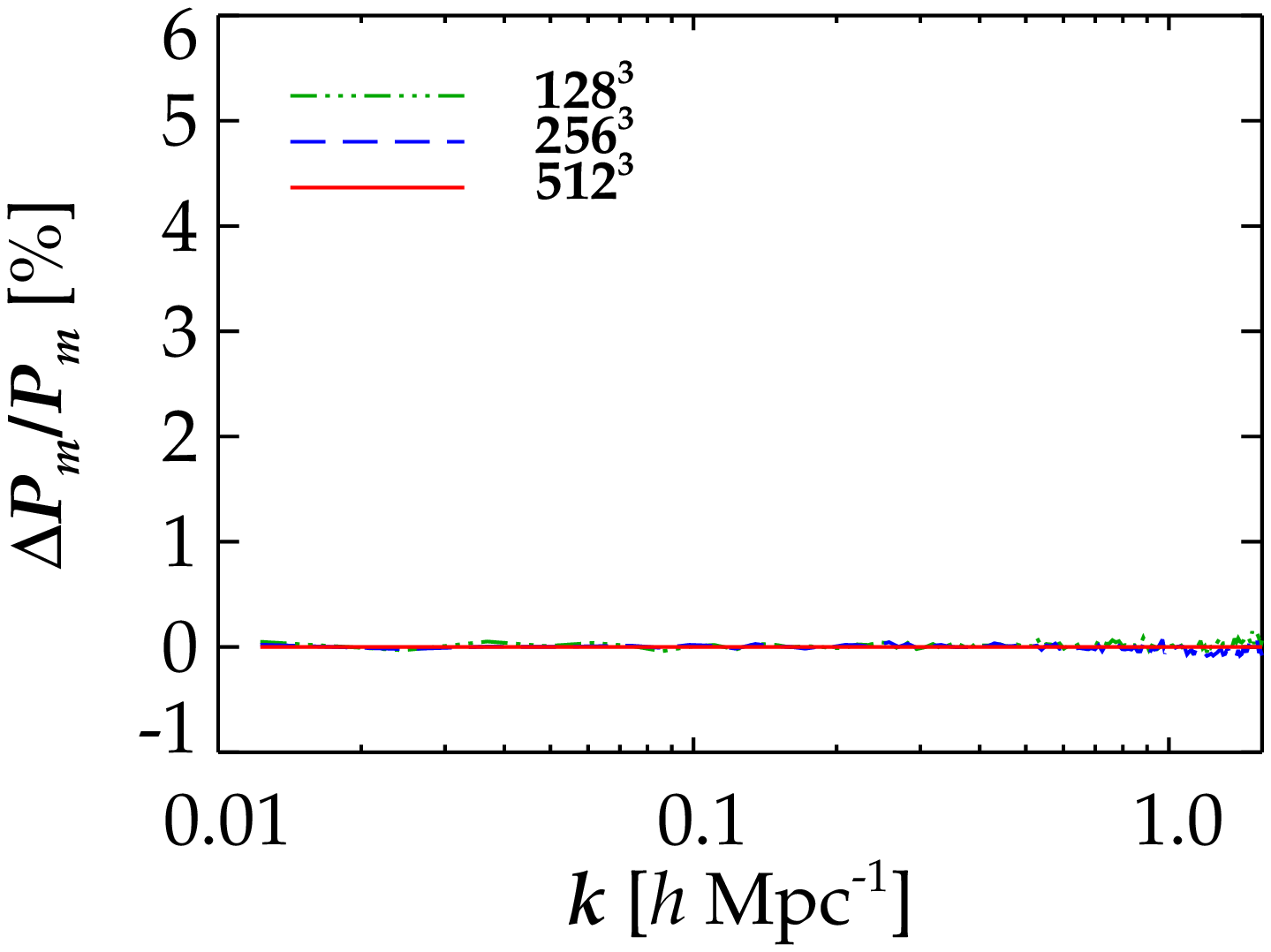}
   \end{minipage}
   \begin{minipage}{0.49\linewidth}
      \hspace*{-0.5cm}\includegraphics[width=1.2\linewidth]{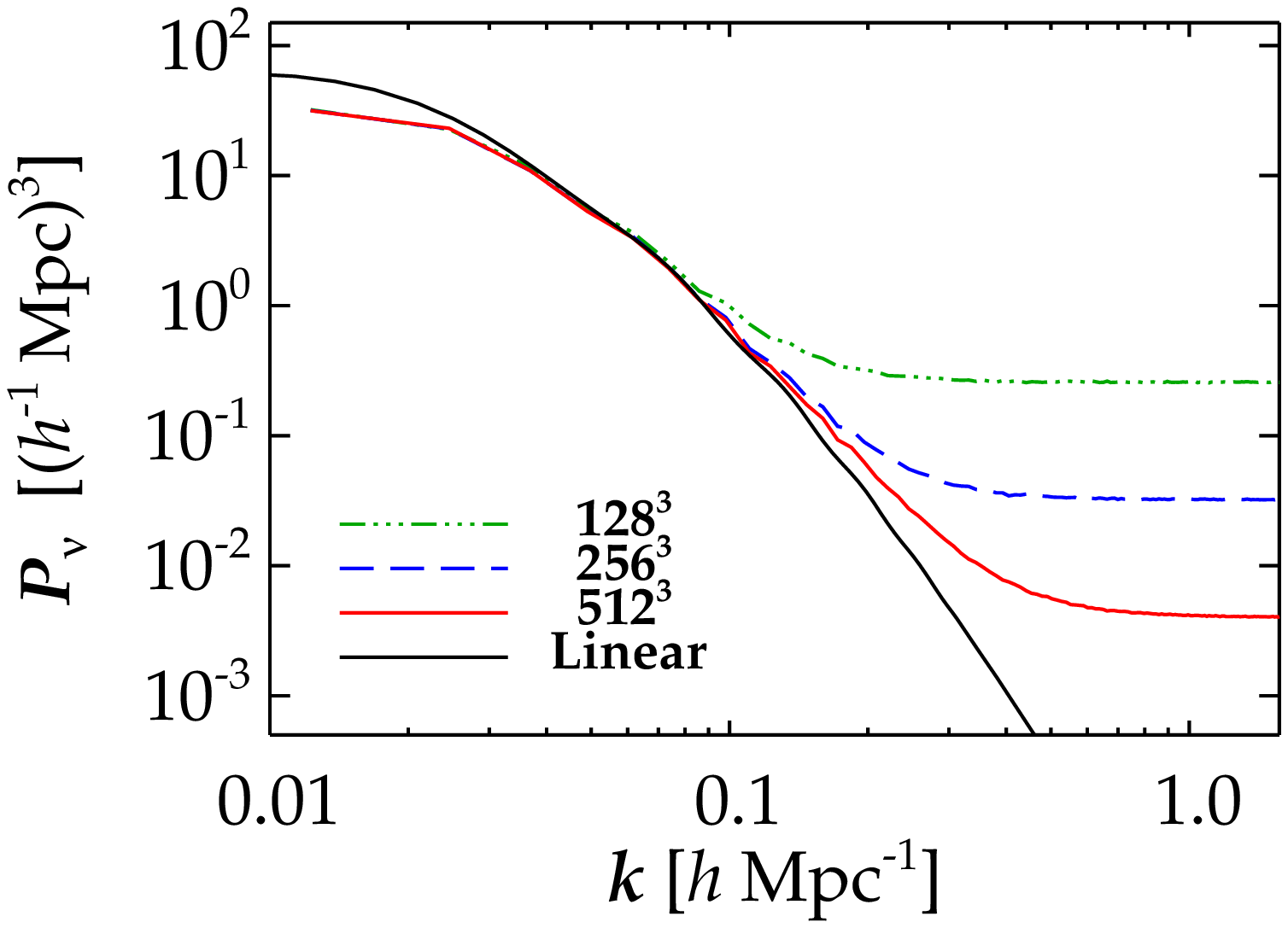}
   \end{minipage}
   \caption{Top left: Differences in $\%$ in the matter power spectra at $z=0$ with
   $\sum m_\nu=0.6$ eV neutrinos and $z_i=4$. The differences are taken with respect to the $1024^3$ neutrino
   simulation. Top right: Neutrino power spectra at $z=0$ with $\sum m_\nu=0.6$ eV neutrinos and $z_i=4$. Bottom left: Differences in $\%$ in the matter power spectra at $z=0$ with $\sum m_\nu=0.3$ eV neutrinos and $z_i=4$. The differences are taken with respect to the $512^3$ neutrino
   simulation. Bottom right: Neutrino power spectra at $z=0$ with $\sum m_\nu=0.3$ eV neutrinos and $z_i=4$.}
   \label{fig:powerall}
\end{figure}

The turnover is not caused by clustering of the neutrino component on small scales. This can be seen by comparing the two figures in the top panel of Fig.~\ref{fig:powerall} (see below for further comments on Fig.~\ref{fig:powerall}). If small-scale neutrino clustering did contribute to the matter power spectrum on scales $k \sim 1 \, h\, {\rm Mpc^{-1}}$ then the matter power spectrum should be affected by the different noise levels seen in the neutrino power spectra. But since the matter power spectrum has converged for $128^3$ to $1024^3$ neutrino $N$-body particles the turnover is not caused by neutrino small-scale clustering.

The characteristic scale at which the turnover appears, $k_{\rm turn}$, moves to smaller $k$ as $\Omega_\nu$ is decreased. $k_{\rm turn}$ is not directly proportional to $\Omega_\nu$ and therefore to the neutrino free-streaming length. Instead, $k_{\rm turn}$ is related to the amount of $\Omega_{\rm CDM}$. Since we keep $\Omega_{m}$ fixed, decreasing $\Omega_\nu$ increases the amount of CDM, and more CDM makes the halos collapse on larger scales. Since the turnover is related to the non-linear collapse of structures, $k_{\rm turn}$ moves to larger scales as $\Omega_\nu$ is decreased. Since there is a linear boundary condition at $k \sim 0.2 h \, {\rm Mpc}^{-1}$, the turnover cannot propagate beyond this value for $\Omega_\nu$ approaching zero.        

The maximum relative magnitude of the power spectrum suppression is roughly given by
\begin{equation}
\left. \frac{\Delta P}{P} \right|_{\rm max} \sim -9.8
\frac{\Omega_\nu}{\Omega_m},
\end{equation}
which is about 20\% larger than the linear theory prediction of
$\Delta P/P|_{\rm lin} \sim -8 \Omega_\nu/\Omega_m$
\cite{Hu:1997mj,Lesgourgues:2006nd}. With future high precision
measurements of the matter power spectrum approaching the 1\%
precision on these scales this effect must be taken into account
even for hierarchical neutrino masses.

The differences for the matter power spectra in the $\sum m_\nu=0.6 \,
{\rm eV}$ case as a function of different numbers of neutrino $N$-body
particles are shown in the top left panel of Fig.~\ref{fig:powerall}. By using different mass
assignment methods, Nearest-Grid-Point, deconvolved Cloud-In-Cell, the
adaptive smoothing kernel of \cite{monaghan} as well as dividing the
simulation volume into equal cubical cells and superposing the
particle distributions for attaining higher spacial resolution, we have tested that the power spectra do not
have any artificial features caused by the particular mass assignment
method used. For all neutrino masses simulated it is necessary to include at least
$128^3$ neutrinos in the $N$-body simulation and give them a thermal velocity to calculate the
matter power spectrum at percent level accuracy on the relevant
scales. For the $\sum m_\nu = 0.6 \, {\rm eV}$ case and $z_i=49$ it is furthermore necessary to include $512^3$ neutrino $N$-body particles to achieve convergence.

The neutrino power spectra for $\sum m_\nu = 0.6 \, {\rm eV}$ as a
function of different numbers of neutrino $N$-body particles are shown
in the top right panel of Fig.~\ref{fig:powerall}. On small scales the neutrino power spectrum is in most cases completely flat because
the coarse sampling of the neutrino velocity distribution introduces
a white noise term. That this is indeed the reason can be seen from
the fact that the noise level decreases rapidly when more neutrino
{\it N}-body particles are used. For runs with $128^3$ to $1024^3$
neutrino {\it N}-body particles the noise effect on the matter power
spectrum is kept well below
the 1\% level on relevant scales. It can also be seen that the neutrino power spectra for
$N_\nu=256^3$ and $N_\nu=512^3$ converge out to $k \simeq 0.2 \, h \, {\rm
Mpc}^{-1}$ and that the $N_\nu=512^3$ and $N_\nu=1024^3$ simulations
converge out to $k \simeq 0.5 \, h \, {\rm Mpc}^{-1}$.

In the bottom panel of Fig.~\ref{fig:powerall} the corresponding convergence of the matter power spectra (left) and the neutrino power spectra (right) for the $\sum m_\nu = 0.3 \, {\rm eV}$ case are shown. The same general trends from the $\sum m_\nu = 0.6 \, {\rm eV}$ case can be seen.

\begin{figure}
   \begin{center}
   \includegraphics[width=0.6\textwidth]{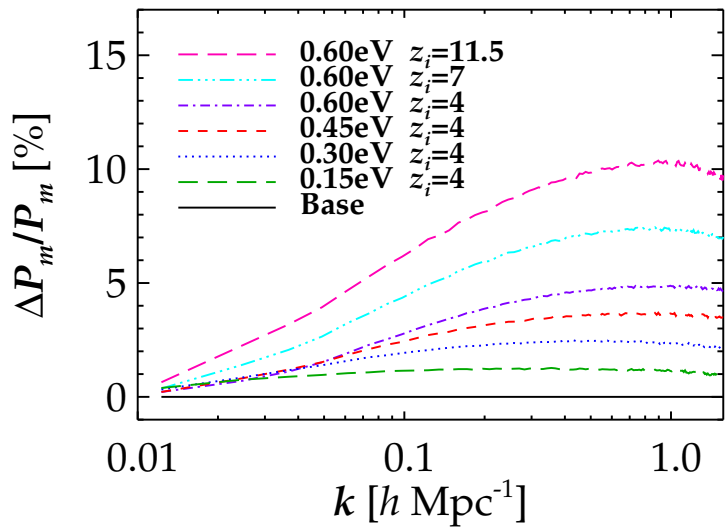}
   \caption{The effect on the matter power spectrum at $z=0$ of neglecting
   thermal velocities in the {\it N}-body simulations.
   The neutrinos have been correctly included in the linear evolution.}
    \label{fig:FD0}
    \end{center}
\end{figure}

\subsection{The Effect of Neglecting the Thermal Component}
Fig.~\ref{fig:FD0} shows the
effect of neglecting the neutrino thermal velocity in the
simulations. In the figure the power spectra are normalized with
respect to the simulation with the highest number of neutrino {\it
N}-body particles for a given mass and starting redshift. As expected,
neglecting the thermal velocity component increases the amplitude
of the power spectrum, because without thermal velocities the neutrinos
act as an extra CDM species, and the effect of free-streaming is neglected once the simulation is started. The effect is
more pronounced at smaller scales, and is largest in the highest
mass neutrino case, since it contributes a larger fraction of
$\Omega_{m}$, even though it has a smaller omitted thermal velocity
component than in the cases of the lower mass neutrinos.

For the case of $\sum m_\nu=0.6 \, {\rm eV}$ the effect is as large
as $\sim$ 5\% for $z_i=4$ increasing to $\sim$ 10\% for $z_i=11.5$
on scales relevant for future large-scale surveys. For higher
starting redshifts the effect is even larger. We
note that for a given starting redshift the effect is proportional
to $\sum m_\nu$ as expected.

\begin{figure}[t]
   \noindent
   \begin{minipage}{0.49\linewidth}
      \hspace*{-1.1cm}\includegraphics[width=1.2\linewidth]{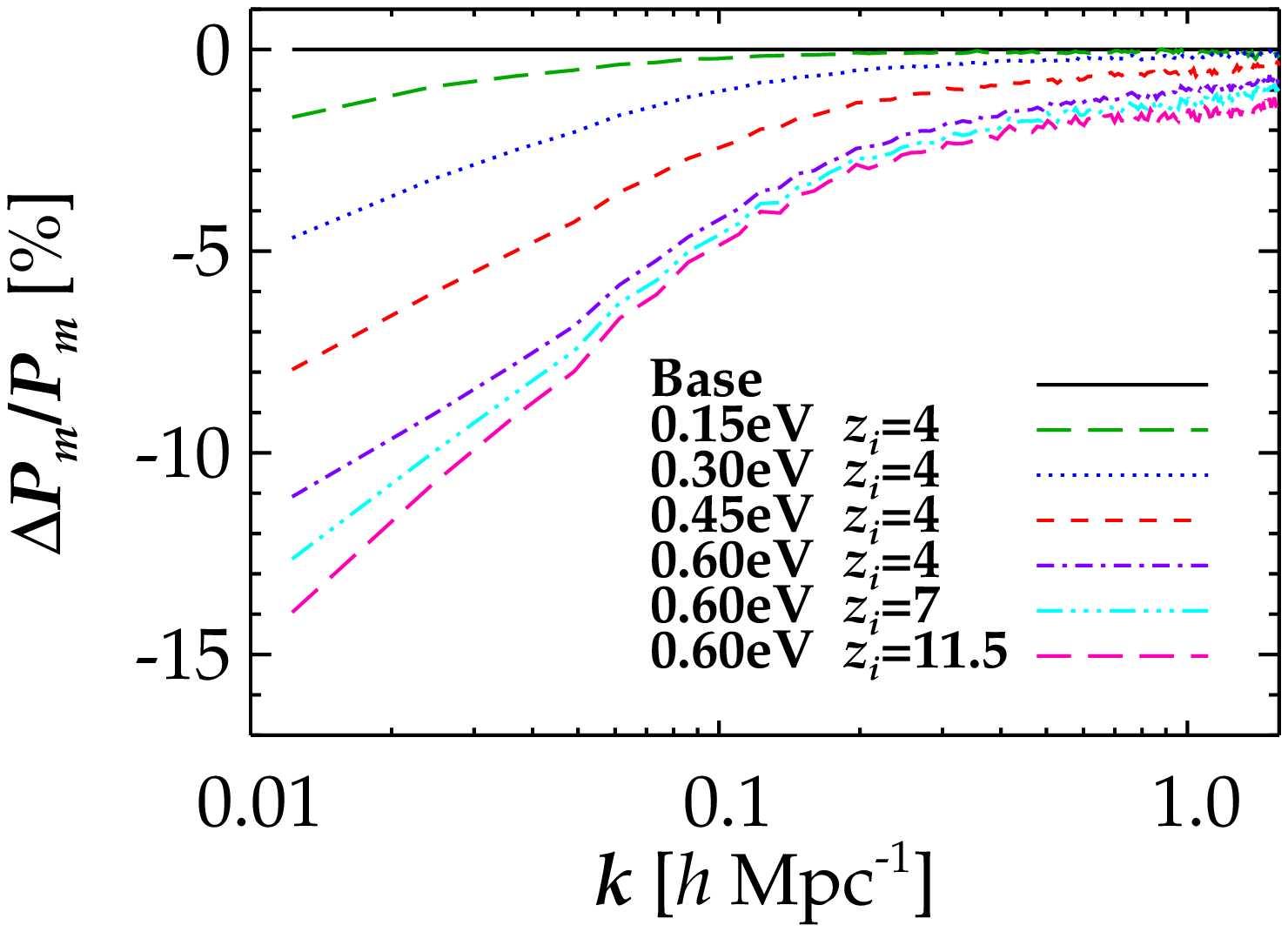}
   \end{minipage}
   \begin{minipage}{0.49\linewidth}
      \hspace*{-0.5cm}\includegraphics[width=1.2\linewidth]{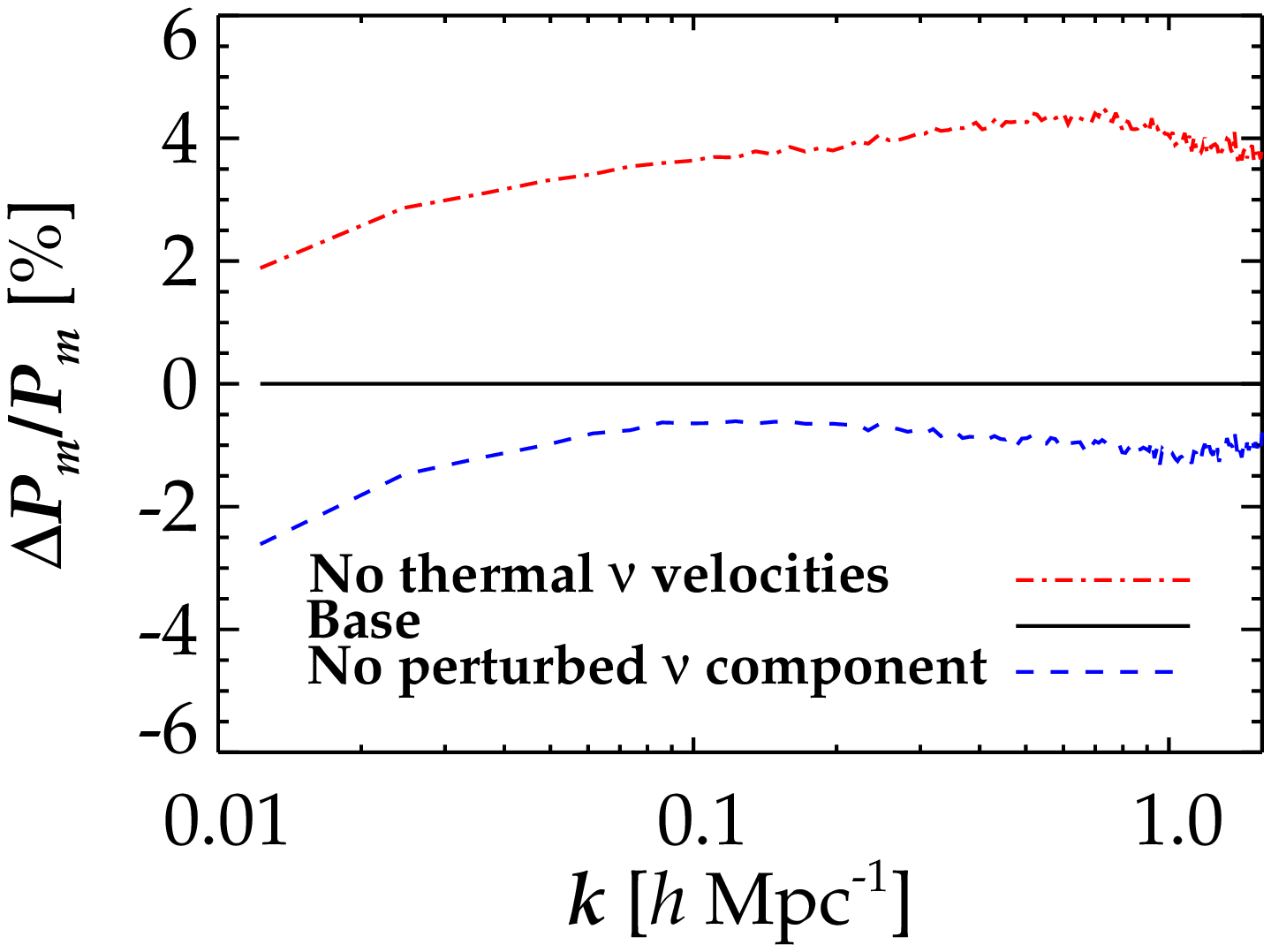}
   \end{minipage}
   \caption{Left: The effect of neglecting the perturbed
    neutrino component at $z=0$ in the {\it N}-body
    simulations. Right: The effect of neglecting the
    thermal velocities or the perturbed
    neutrino component for $\sum m_\nu
    = 0.15 \, {\rm eV}$ and a starting redshift of 49. The 'Base' has
    been estimated, as explained in the text. In all cases the
    neutrinos have been correctly included in the linear evolution.}
    \label{fig:nu0}
\end{figure}

\subsection{High $z_i$ Low $m_\nu$ Approximation}
Including neutrinos with a total mass in the range $0.3 \sim 0.6 \, {\rm
  eV}$ in {\it N}-body simulations can be done consistently even at
  redshifts required for getting the absolute power spectrum
  accurately. However, for very light neutrinos, $\sum m_\nu \sim 0.15 \, {\rm eV}$,
  the thermal velocities are semi-relativistic at $z=49$. This
  will not only render Newtonian dynamics inaccurately but will also,
  depending on the number of neutrino {\it N}-body particles used,
  erase the initial neutrino large-scale structure since typical neutrino
  gravitational flow velocities are of order $\sim 5 \, {\rm km \, s^{-1} }$,
much smaller than the thermal velocities of individual particles. Furthermore,
the high neutrino velocities make the {\it N}-body simulation timestep
very short, increasing simulation times substantially.

For these reasons it is desirable to develop an approximate method
which can be used to calculate the matter power spectrum at the 1\%
level without including thermal velocities. In Fig.~\ref{fig:nu0}
(left) we show the effect of neglecting the perturbed neutrino
component in the {\it N}-body simulations. The CDM $N$-body particle
mass is still scaled according to the total matter density. As
expected this gives less power because of the lack of feed-back from
the large-scale modes of the neutrino component to the CDM
component. By increasing $z_i$ the neutrino perturbations are omitted
over a longer time span and therefore affect and decrease the power
spectrum more.

It is important to notice, that this lack of power is very small in
the $\sum m_\nu = 0.15 \, {\rm eV}$ case for $k > 0.1 \, h\, {\rm
Mpc^{-1}}$, i.e.\ on scales where non-linear corrections become
important.

Ideally, the ratio of a pure $\Lambda$CDM power spectrum to that of a
power spectrum with a given neutrino mass,  $P_{\Lambda {\rm CDM}}/P_{\nu
  \Lambda {\rm CDM}}$, should be independent of $N$-body starting redshift, as long as $z_i$ is not too low.
The presence of non-linearities at the time when the ICs are
calculated make this power spectrum ratio marginally dependent
on $z_i$, which is also visible in Fig.~\ref{fig:damping}
in the $\sum m_\nu = 0.6 \, {\rm eV}$ case.

Now, focusing on the $\sum m_\nu = 0.15 \, {\rm eV}$ case and $z_i=49$, we would
like to estimate the error made by not including the perturbed
neutrino component in the $N$-body simulation. Because the damping
of the power spectrum is nearly independent of $z_i$ it is possible
to estimate the matter power spectrum with neutrinos and thermal
velocities without actually performing the simulation. More specifically, assuming that the damping effect is independent of $z_i$, the matter power spectrum at $z=0$ for an $N$-body simulation started at $z_i=49$ with neutrinos and thermal velocities can be found from $P_{\Lambda {\rm CDM}}(z_i=4)/P_{\nu \Lambda {\rm CDM}}(z_i=4) = P_{\Lambda {\rm CDM}}(z_i=49)/P_{\nu \Lambda {\rm CDM}}(z_i=49)$. This estimated power spectrum is used as the 'Base' in the right panel of Fig.~\ref{fig:nu0}. This figure shows the error made by neglecting the
perturbed neutrino component. On scales relevant for $N$-body
simulations, i.e. $k > 0.1 \, h\, {\rm Mpc^{-1}}$, the error made is
at the 1\% level. For comparison the effect of neglecting the
thermal velocities is also shown, and it can be seen that it is a
better approximation to neglect the perturbed neutrino component
than to include it but neglect the thermal velocity component.

In the $\sum m_\nu = 0.6 \, {\rm eV}$ case we have all the relevant power spectra at
starting redshifts of 4 and 49. Therefore we can test the validity
of estimating the power spectrum with neutrinos and thermal
velocities. We have found that the 'Base' in the right panel of Fig.~\ref{fig:nu0} is overestimated so that the error made by neglecting the perturbed
neutrino component is an upper bound.

We have tried to decrease the error made by neglecting the perturbed
neutrino component. This has been done by including the perturbed neutrino component
without thermal velocities in the $N$-body simulation, but switching
off the effect of the gravitational field on the neutrinos, so that
the gravitational flow velocities of the neutrinos at $z_i$ are
frozen in time. This should improve the estimate of the matter power
spectrum in two ways: First, by including the
perturbed neutrino component the power spectrum will increase. Second, by switching off gravity, artificial
small-scale structures will not form. With this method the error
was decreased, especially on large scales in the $\sum m_\nu = 0.6 \, {\rm eV}$
case for $z_i=4$, but the error did not approach percent level
precision. On the contrary, in the $\sum m_\nu = 0.15 \, {\rm eV}$ case and for
$z_i=49$ the error $increased$ to the 10-15\% level. The
reason is that at such a high redshift the neutrino gravitational
flow velocities and density perturbations are very small effectively
freezing the neutrinos on a regular grid. Because of neutrino
$N$-body particle shot-noise this stationary regular grid deflects
the CDM particles and therefore reduces the amount of structure
forming.

\section{Discussion and Conclusions}\label{secDiscussion}

We have performed a precise calculation of the effect of including
neutrino dark matter on the matter power spectrum. The effect of
thermal neutrino motion has been included directly, and we have shown that this effect
changes the matter power spectrum significantly.

Specifically, we find that the suppression of power due to the
presence of massive neutrinos is increased by non-linear effects.
Whereas in linear theory the suppression of power on small scales is
given roughly by $\Delta P/P \sim -8 \Omega_\nu/\Omega_m$, the full
non-linear calculation gives $\Delta P/P \sim -9.8
\Omega_\nu/\Omega_m$ at a scale of $k \sim 0.5-1 \, h \, {\rm
Mpc}^{-1}$, i.e.\ an increase of about 20\%.

On smaller scales the non-linear contribution to the suppression
decreases again. This effect has previously been noted in
semi-analytic studies such as \cite{Hannestad:2005bt}, and occurs on
a scale which depends upon the amount of CDM and the neutrino free-streaming length.

The increased suppression due to non-linear effects is highly
relevant for future high precision large-scale structure and weak
lensing surveys. Even for neutrino masses approaching the lower
bound, found from oscillation experiments, it is large enough to bias the
estimate of other cosmological parameters. Conversely, it provides a very
distinct signature which could allow for the detection and
measurement of even very low-mass neutrinos.

For $k > 0.1 \, h\, {\rm Mpc^{-1}}$, which include the modes for which
{\it N}-body simulations are needed, it is a better approximation to
the ``true'' power spectrum to neglect the perturbed neutrino
component than to include it without thermal velocities. In both
cases the approximation to the ``true'' power spectrum is better
than if a standard $\Lambda$CDM model was assumed.

But only in our lowest mass neutrino case, $\sum m_\nu = 0.15 \, {\rm eV}$, is
the error on the non-linear matter power spectrum made by neglecting
the perturbed neutrino component in the $N$-body simulation at the
desired 1\% level for the relevant modes, even for an $N$-body
simulation with a starting redshift as high as 49. The error would be
decreased further for smaller neutrino masses.

An alternative method for implementing the physics of neutrinos in
$N$-body simulations is to represent the neutrino component as a fluid
and solve the corresponding fluid equations on a grid. This method
would be particularly useful for neutrino masses close to the lower
observational bound.


\section*{Acknowledgements}
We acknowledge computing resources from the Danish Center for
Scientific Computing (DCSC). We thank Yvonne Wong for discussions
and comments.



\section*{References} 



\begin{thebibliography}{00}

\bibitem{Fogli:2006yq}
  G.~L.~Fogli {\it et al.},
  Phys.\ Rev.\  D {\bf 75}, 053001 (2007)
  [arXiv:hep-ph/0608060].

\bibitem{Maltoni:2004ei}
  M.~Maltoni, T.~Schwetz, M.~A.~Tortola and J.~W.~F.~Valle,
  New J.\ Phys.\  {\bf 6}, 122 (2004)
  [arXiv:hep-ph/0405172].

\bibitem{Zunckel:2006mt}
  C.~Zunckel and P.~G.~Ferreira,
  [arXiv:astro-ph/0610597].

\bibitem{Cirelli:2006kt}
  M.~Cirelli and A.~Strumia,
  JCAP {\bf 0612} (2006) 013
  [arXiv:astro-ph/0607086].

\bibitem{Goobar:2006xz}
  A.~Goobar, S.~Hannestad, E.~Mortsell and H.~Tu,
  JCAP {\bf 0606} (2006) 019
  [arXiv:astro-ph/0602155].

\bibitem{Kristiansen:2006xu}
  J.~R.~Kristiansen, H.~K.~Eriksen and O.~Elgaroy,
  Phys.\ Rev.\  D {\bf 74}, 123005 (2006).

\bibitem{Seljak:2006bg}
  U.~Seljak, A.~Slosar and P.~McDonald,
  JCAP {\bf 0610}, 014 (2006)
  [arXiv:astro-ph/0604335].

\bibitem{Hannestad:2003xv}
  S.~Hannestad,
  JCAP {\bf 0305}, 004 (2003)
  [arXiv:astro-ph/0303076].

\bibitem{Hannestad:2006zg}
  S.~Hannestad,
  Ann.\ Rev.\ Nucl.\ Part.\ Sci.\  {\bf 56} (2006) 137
  [arXiv:hep-ph/0602058].

\bibitem{Hannestad:2007tu}
  S.~Hannestad,
  arXiv:0710.1952 [hep-ph].

\bibitem{Hannestad:2007dd}
  S.~Hannestad, A.~Mirizzi, G.~G.~Raffelt and Y.~Y.~Y.~Wong,
  JCAP {\bf 0708}, 015 (2007)
  [arXiv:0706.4198 [astro-ph]].

\bibitem{LSST}
www.lsst.org.

\bibitem{Klypin:1992sf}
  A.~Klypin, J.~Holtzman, J.~Primack and E.~Regos,
  Astrophys.\ J.\  {\bf 416}, 1 (1993)
  [arXiv:astro-ph/9305011].

\bibitem{Primack:1994pe}
  J.~R.~Primack, J.~Holtzman, A.~Klypin and D.~O.~Caldwell,
  Phys.\ Rev.\ Lett.\  {\bf 74}, 2160 (1995)
  [arXiv:astro-ph/9411020].

\bibitem{Colin:2007bk}
  P.~Colin, O.~Valenzuela and V.~Avila-Reese,
  [arXiv:0709.4027 [astro-ph]].

\bibitem{Singh:2002de}
  S.~Singh and C.~P.~Ma,
  Phys.\ Rev.\  D {\bf 67}, 023506 (2003)
  [arXiv:astro-ph/0208419].

\bibitem{Ringwald:2004np}
  A.~Ringwald and Y.~Y.~Y.~Wong,
  JCAP {\bf 0412}, 005 (2004)
  [arXiv:hep-ph/0408241].


\bibitem{Ma}
  C.~P.~Ma and E.~Bertschinger,
  Astrophys.\ J.\  {\bf 455}, 7 (1995)
  [arXiv:astro-ph/9506072].

\bibitem{Scoccimarro1}
  R.~Scoccimarro,
  Mon.\ Not.\ Roy.\ Astron.\ Soc.\  {\bf 299}, 1097 (1998)
  [arXiv:astro-ph/9711187].

\bibitem{Bouchet}
  F.~R.~Bouchet, S.~Colombi, E.~Hivon and R.~Juszkiewicz,
  Astron.\ Astrophys.\  {\bf 296}, 575 (1995)
  [arXiv:astro-ph/9406013].

\bibitem{Springel2}
  V.~Springel,
  Mon.\ Not.\ Roy.\ Astron.\ Soc.\  {\bf 364}, 1105 (2005)
  [arXiv:astro-ph/0505010].

\bibitem{CAMB}
  A.~Lewis and S.~Bridle,
  Phys.\ Rev.\ D {\bf 66}, 103511 (2002)
  [arXiv:astro-ph/0205436].

\bibitem{CMBFAST}
  U.~Seljak and M.~Zaldarriaga,
  Astrophys.\ J.\  {\bf 469}, 437 (1996)
  [arXiv:astro-ph/9603033].

\bibitem{Crocce}
  M.~Crocce, S.~Pueblas and R.~Scoccimarro,
  Mon.\ Not.\ Roy.\ Astron.\ Soc.\  {\bf 373}, 369 (2006)
  [arXiv:astro-ph/0606505].

 \bibitem{monaghan}J.~J.~Monaghan and J.~C.~Lattanzio, Astron. Astrophys. {\bf 149}, 135 (1985).


\bibitem{spergel1}
  D.~N.~Spergel {\it et al.}  [WMAP Collaboration],
  Astrophys.\ J.\ Suppl.\  {\bf 170}, 377 (2007)
  [arXiv:astro-ph/0603449].

\bibitem{Hannestad:2006as}
  S.~Hannestad, H.~Tu and Y.~Y.~Y.~Wong,
  JCAP {\bf 0606}, 025 (2006)
  [arXiv:astro-ph/0603019].

\bibitem{saito1}
  S.~Saito, M.~Takada and A.~Taruya,
  [arXiv:0801.0607 [astro-ph]].




%



%

\bibitem{Zeldovich:1969sb}
  Y.~B.~Zeldovich,
  Astron.\ Astrophys.\  {\bf 5}, 84 (1970).

\bibitem{Dicus:1982bz}
D.~A.~Dicus, E.~W.~Kolb, A.~M.~Gleeson, E.~C.~Sudarshan,
V.~L.~Teplitz and M.~S.~Turner,
Phys.\ Rev.\ D {\bf 26}, 2694 (1982).


\bibitem{Dodelson:1992km}
S.~Dodelson and M.~S.~Turner,
Phys.\ Rev.\ D {\bf 46}, 3372 (1992).

\bibitem{Hannestad:1995rs}
S.~Hannestad and J.~Madsen,
Phys.\ Rev.\ D {\bf 52}, 1764 (1995) [arXiv:astro-ph/9506015].


\bibitem{Dolgov:1997mb}
A.~D.~Dolgov, S.~H.~Hansen and D.~V.~Semikoz,
Nucl.\ Phys.\ B {\bf 503}, 426 (1997) [arXiv:hep-ph/9703315].


\bibitem{Steigman:2001px}
G.~Steigman,
arXiv:astro-ph/0108148.

\bibitem{Mangano:2001iu}
G.~Mangano, G.~Miele, S.~Pastor and M.~Peloso,
[arXiv:astro-ph/0111408].

\bibitem{osc}S.~Hannestad, Physical Review D {\bf 65}, 083006
(2002).

\bibitem{Mangano:2005cc}
  G.~Mangano, G.~Miele, S.~Pastor, T.~Pinto, O.~Pisanti and P.~D.~Serpico,
  Nucl.\ Phys.\ B {\bf 729}, 221 (2005)
  [arXiv:hep-ph/0506164].






%
\bibitem{Hu:1997mj}
  W.~Hu, D.~J.~Eisenstein and M.~Tegmark,
  Phys.\ Rev.\ Lett.\  {\bf 80}, 5255 (1998)
  [arXiv:astro-ph/9712057].

\bibitem{Lesgourgues:2006nd}
  J.~Lesgourgues and S.~Pastor,
  Phys.\ Rept.\  {\bf 429}, 307 (2006)
  [arXiv:astro-ph/0603494].

\bibitem{Hannestad:2005bt}
  S.~Hannestad, A.~Ringwald, H.~Tu and Y.~Y.~Y.~Wong,
  JCAP {\bf 0509}, 014 (2005)
  [arXiv:astro-ph/0507544].

\end{thebibliography}
\end{document}